\documentstyle[prd,aps,eqsecnum,preprint,tighten,psfig,floats]{revtex}
\begin{document}
\draft
\preprint{FSU-HEP-970411}

\title{Photoproduction of jets at HERA in next-to-leading order QCD}

\author{B.W. Harris and J.F. Owens}

\address{Physics Department \\ Florida State University \\
Tallahassee, Florida 32306-3016, USA}

\date{April 1997}

\maketitle

\begin{abstract}

A new next-to-leading order Monte Carlo program for 
the calculation of jet cross sections in photoproduction 
is described.  The contributions from both resolved and direct components 
are included to ${\cal O}(\alpha \alpha_s^2)$.  Properties of the predictions 
for various inclusive jet and dijet observables are discussed and  
comparisons with HERA data are presented.

\end{abstract}


\section{Introduction}

Electromagnetic interactions have long been used to study both hadronic 
structure and strong interaction dynamics. Examples include deep inelastic 
lepton-nucleon scattering, hadronic production of lepton pairs, the 
production of photons with large transverse momenta, and various
photoproduction processes involving the scattering of real or very low mass 
virtual photons from hadrons. In particular, the photoproduction of jets with 
large transverse momenta is calculable in QCD and offers additional 
complementary information to that obtained from the study of the
hadroproduction of jets \cite{owens}. In the photoproduction case there are
contributions where the photon's energy contributes entirely to the 
hard-scattering subprocess; these are often referred to collectively as the 
direct component. In addition, a real photon can interact via its hadronic 
substructure. These contributions comprise the resolved component, a review of
which is contained in Ref.\cite{drees}. Therefore, the photoproduction of jets
allows one to investigate new production mechanisms, probe the hadronic
substructure of the photon, and study the conventional hadronic jet 
production mechanisms which these processes have in common with the
hadroproduction case.

Two basic approaches are commonly employed for generating predictions for
hard-scattering processes. On the one hand, it is possible to perform 
calculations for a 
specific observable in which the integrals over the subprocess variables are 
done analytically, leaving only the convolutions with the parton distributions
to be done numerically. On the other hand, if the subprocess integrations and
the parton distribution convolutions are done using Monte Carlo techniques, it 
is possible to generate predictions for a variety of different observables
simultaneously. This latter approach is sometimes referred to as a fully
differential Monte Carlo calculation. Several groups have performed 
next-to-leading-order calculations of jet 
photoproduction in varying degrees of generality using one or the other of
these two approaches.  
In \cite{boo}, \cite{bod}, and \cite{klas1} subprocesses which involved 
the photon were kept up to ${\cal O}(\alpha \alpha_s^2)$.  
Thus, the direct component
was calculated to next-to-leading-logarithm accuracy while the resolved 
component was calculated in the leading-logarithm approximation.
The resolved component was calculated to next-to-leading-logarithm accuracy 
in \cite{gsjet}, \cite{greco}, \cite{au1}, and \cite{bod2} for 
single inclusive production.  For a review see \cite{kramer}.  
Recently, both direct and resolved contributions 
calculated to next-to-leading-logarithm accuracy have begun to appear 
\cite{ho}, \cite{kl}, \cite{au2} in a fully 
differential form.

The purpose of this work is to present a discussion of a calculation 
which is based on the phase space slicing method using two cutoffs 
\cite{bergmann}.  Both the direct and resolved components are included 
to next-to-leading-logarithm accuracy.  The result is fully differential and 
implemented in a Monte Carlo 
style program which allows the simultaneous histogramming of many 
distributions incorporating experimental cuts.  It represents an improvement 
of earlier results which included the direct component only at NLO 
\cite{boo} and an elaboration of the very brief results for both 
components already presented \cite{ho}.  Details of the 
calculational method are presented as well as comparisons with recent data. 
Some comments on various unsettled issues that arise when comparing 
with dijet cross sections are also given.

The remainder of the paper is as follows.  The phase space slicing method 
is reviewed in Sec.\ II.  Numerical results 
are compared with H1 and ZEUS data and related physics issues are discussed 
in Sec.\ III\@, while the conclusions are given in Sec.\ IV\@.

\section{Method}

In this section we describe the calculation of QCD corrections to 
two-jet production in electron-proton scattering using the phase space 
slicing method.  Before discussing the QCD corrections, it is necessary 
to recount the connection 
between electron-proton and photon-proton scattering.  For 
small photon virtualities $Q^2$ the two are related using the 
Weizs\"{a}cker-Williams approximation \cite{ww} wherein one assumes  
that the incoming electron beam is equivalent to a broad-band photon beam.  
The cross section for electron-proton scattering is then given as a 
convolution of the photon distribution in an electron and the 
photon-proton cross section
\begin{equation}
d \sigma (ep \rightarrow eX) = \int_{y_{\rm min}}^{y_{\rm max}} dy 
F_{\gamma/e}(y) \, d \sigma (\gamma p \rightarrow X ) .
\end{equation}
The integration limits and the maximum photon virtuality $Q^2_{\rm max}$ 
are determined from the (anti) tagging conditions of the experiment.  
The energy of the photon is related to the energy of the 
incident electron by $E_{\gamma}=yE_e$.
We used the improved photon 
distribution in an electron \cite{fmnr} given by the formula 
\begin{equation}
F_{\gamma/e}(y) = \frac{\alpha_{\rm em}}{2 \pi} \left\{ \frac{1+(1-y)^2}{y} 
\ln \frac{Q^2_{\rm max}(1-y)}{m^2_e y^2}+2m^2_e y \left[ 
\frac{1}{Q^2_{\rm max}} - \frac{(1-y)}{m^2_e y^2} \right] \right\}
\end{equation}
where $m_e$ is the electron mass and $\alpha_{\rm em}$ is the electromagnetic 
coupling.  Within this approximation, QCD corrections to electron-proton 
scattering correspond to QCD corrections to photon-proton scattering 
to which we now turn.

In this version of the phase space slicing method \cite{bergmann} 
two small cutoffs $\delta_s$ and $\delta_c$ are used to delineate 
regions of phase space where soft and collinear singularities occur.
Let the four vectors of the three-body subprocesses be labeled 
$p_1+p_2 \rightarrow p_3+p_4+p_5$, and define the Mandelstam 
invariants $s_{ij}=(p_i+p_j)^2$ and $t_{ij}=(p_i-p_j)^2$.  
Consider, for example, the 
$\gamma(p_1) g(p_2) \rightarrow q(p_3) {\overline q}(p_4) g(p_5)$ 
subprocess whose matrix element becomes singular as the energy $E_5$ 
of the final state gluon becomes soft.  
Define the soft region $S$ by 
$0 < E_5 < \delta_s \sqrt{s_{12}}/2$ and the complementary hard region 
$H$ by $\delta_s \sqrt{s_{12}}/2 < E_5 < \sqrt{s_{12}}/2$, both in the 
$p_1+p_2$ rest frame.
The two-to-three body contribution to the cross section is then 
decomposed as
\begin{eqnarray}
\sigma & = & \frac{1}{2s_{12}} \int |M|^2 d\Gamma_3 \nonumber \\
       & = & \frac{1}{2s_{12}} \int_{S} |M|^2 d\Gamma_3 + 
             \frac{1}{2s_{12}} \int_{H} |M|^2 d\Gamma_3
\end{eqnarray}
where $|M|^2$ is the three-body squared matrix element and $d\Gamma_3$ 
is the three-body phase space.
Within $S$ one sets $p_5=0$ ${\em everywhere}$ except in the 
denominators of the matrix elements and then analytically integrates 
over the unobserved degrees of freedom in $n$ space-time dimensions.  
The result, proportional to the leading order cross section,  
contains double and single poles in $n-4$, and double and single 
logarithms in the soft cutoff $\delta_s$.  
Terms of order $\delta_s$ are neglected.
Next, the collinear regions of phase space are defined to be those where 
any invariant ($s_{ij}$ or $t_{ij}$) becomes smaller in magnitude than 
$\delta_c s_{12}$.  The hard region is then decomposed into collinear 
$C$, and non-collinear $\overline{C}$, regions as follows:
\begin{equation}
\frac{1}{2s_{12}} \int_{H} |M|^2 d\Gamma_3 = 
\frac{1}{2s_{12}} \int_{HC} |M|^2 d\Gamma_3 + 
\frac{1}{2s_{12}} \int_{H\overline{C}} |M|^2 d\Gamma_3. 
\end{equation}
Within $HC$ one retains only the leading pole of the vanishing invariant 
in the squared matrix elements.  Exact collinear kinematics 
are used to define the integration domain of $HC$, which is valid so long as 
$\delta_c \ll \delta_s$.  The integrations over the unobserved degrees of 
freedom are then performed analytically in $n$ space-time dimensions giving a 
factorized result where single poles in $n-4$, and single logarithms in both 
cutoffs $\delta_c$ and $\delta_s$, multiply splitting functions 
and lower-order squared matrix elements.  
Terms of order $\delta_c$ and $\delta_s$ are neglected.
The soft and final state hard collinear singularities cancel upon
addition of the interference of the leading order diagrams with the 
renormalized one-loop virtual diagrams.  The remaining initial state
collinear singularities are factorized and absorbed into the parton 
distributions.
The integrations over the singularity-free portion of the three-body
phase space $H\overline{C}$ are performed using standard Monte Carlo 
techniques.
When all of the contributions are combined at the histogramming stage  
the cutoff dependences cancel, provided one looks at a suitably defined 
infrared-safe observable.  An example will be given below.

The matrix elements squared for all two-to-two and two-to-three 
parton-parton scattering subprocesses through ${\cal O}(\alpha_s^3)$ 
are from the paper of Ellis and Sexton \cite{es}, and those for the 
photon-parton scattering subprocesses through ${\cal O}(\alpha \alpha_s^2)$  
are from the paper of Aurenche {\it et al}.\ \cite{aur}.

In order to compare with experimental data, a suitable jet definition must 
be chosen.  For this study, the Snowmass algorithm \cite{snow} 
has been used, supplemented with an $R_{\rm sep}$ constraint \cite{eks} as 
follows.  Let $\eta_i$ ($\eta_j$) be the pseudorapidity and $\phi_i$ 
($\phi_j$) be the azimuthal angle of parton $i$ ($j$) and then define
\begin{equation}
R_{ij} = \left[ (\eta_i-\eta_j)^2+(\phi_i-\phi_j)^2\right]^{1/2}.
\end{equation}
Partons $i$ and $j$ are merged if
\begin{equation}
R_{ij} \leq {\rm min} \left[ 
\frac{E_{T_i}+E_{T_j}}{{\rm max}(E_{T_i},E_{T_j})} R, R_{\rm sep} \right].
\end{equation}
Note that $R_{\rm sep}=2R$ corresponds to having no $R_{\rm sep}$ constraint.
The 4-vector recombination is of the type
\begin{eqnarray}
E_T^{\rm jet} & = & E_T^i + E_T^j \nonumber \\
\eta^{\rm jet} & = & (\eta^i E_T^i + \eta^j E_T^j)/E_T^{\rm jet} 
\nonumber \\
\phi^{\rm jet} & = & (\phi^i E_T^i + \phi^j E_T^j)/E_T^{\rm jet}.
\end{eqnarray}
Other types of jet algorithms may also be used.

\section{Results}

\subsection{General properties}

\begin{figure}
\centerline{\hbox{\psfig{figure=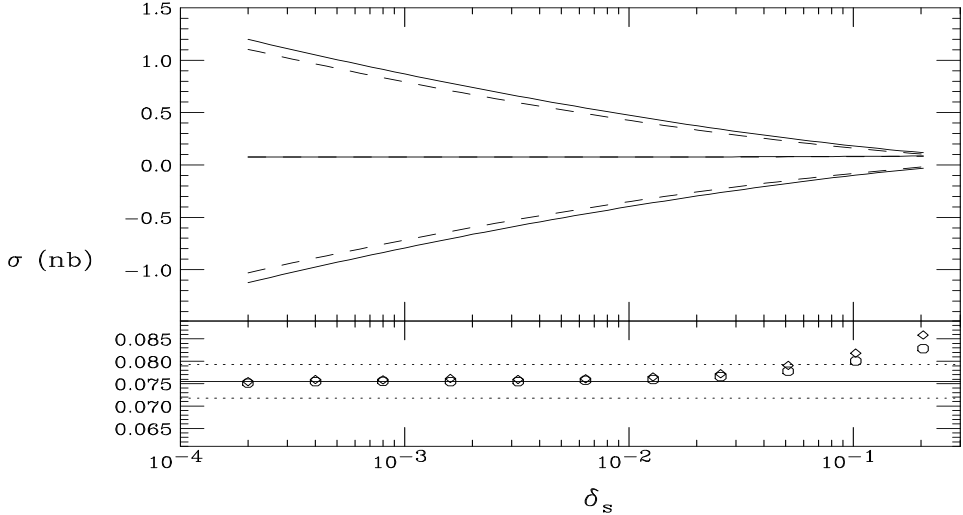,width=5.0in,height=3.09in}}}
\caption{The single jet inclusive cross section integrated over 
$30\, {\rm GeV} < E_T^{\rm jet} < 40\, {\rm GeV}$ and $0<\eta^{\rm jet}<2$.  
The two-body (negative) and three-body (positive) contributions together 
with their sum are shown as a function 
of $\delta_s$ with $\delta_c=\delta_s / a$ for $a=100$ (dash line) and 
$a=200$ (solid line). The bottom enlargement shows the sum for $a=100$ 
(circles) and $a=200$ (diamonds) relative to $\pm 5\%$ (dot line) of the 
average of the last four points (solid line).}
\end{figure}
Using the results of the method described in the previous section we 
have constructed a program to calculate jet photoproduction cross sections.  
The program uses Monte Carlo integration so it is 
possible to implement experimental cuts, provided that they are 
defined in terms of partonic variables.

The CTEQ4 \cite{cteq4} proton parton distribution set was used 
throughout, in conjunction with either the GRV \cite{grv} or 
GS \cite{gs} photon set.  Unless otherwise noted, 
the GRV set was used.  The calculation 
was performed in the ${\overline {\rm MS}}$ factorization scheme. 
Therefore, the corresponding ${\overline {\rm MS}}$ distribution sets 
were used when working at next-to-leading order (NLO).  For leading order (LO) 
results, the appropriate LO sets were used.  
Similarly, the two-loop version of the strong coupling $\alpha_s$ was 
used with matching across quark thresholds for the NLO results, and at 
LO the one-loop value was used.  The value of $\Lambda^{{\rm QCD}}$
was taken from the proton parton distribution set.

For future use we define three sets of beam energy and tagging 
conditions as follows: 1993 H1 where 
$E_e=26.7\, {\rm GeV}$, $E_p=820\, {\rm GeV}$, 
$Q^2_{\rm max}=0.01\, {\rm GeV}^2$ 
and $0.25<y<0.7$, 1993 ZEUS where $E_e=26.7\, {\rm GeV}$, 
$E_p=820\, {\rm GeV}$, 
$Q^2_{\rm max}=4\, {\rm GeV}^2$ and $0.2<y<0.85$, and finally 
1994 ZEUS where $E_e=27.5\, {\rm GeV}$, $E_p=820\, {\rm GeV}$, 
$Q^2_{\rm max}=4\, {\rm GeV}^2$ and $0.25<y<0.8$.

As explained above, 
when contributions from the two-body and three-body pieces are combined 
to form a suitably defined infrared-safe observable the cutoff 
dependences cancel.  This provides a check on the method and also on  
its implementation.  
As stated in Sec.\ II, the method 
requires that $\delta_c \ll \delta_s$.  Therefore, we take 
$\delta_c=\delta_s / a$ with $a \sim {\cal O}(100)$ and verify that the 
result is independent of $a$.
As an example, the single jet inclusive cross section integrated over 
$30\, {\rm GeV} < E_T^{\rm jet} < 40\, {\rm GeV}$ and $0<\eta^{\rm jet}<2$ 
is shown in FIG.\ 1 for the 1994 ZEUS conditions with $R=1$ and 
$R_{\rm sep}=2$.  The two-body (negative) and three-body (positive) 
contributions together with their sum are shown as a function 
of $\delta_s$ for $a=100$ (dash line) and $a=200$ (solid line).   
The bottom enlargement shows the sum for $a=100$ 
(circles) and $a=200$ (diamonds) relative to $\pm 5\%$ (dot line) of the 
actual result (solid line).  For this particular observable, the result 
is sufficiently independent of the cutoffs below $\delta_s=10^{-2}$.

\begin{figure}
\centerline{\hbox{\psfig{figure=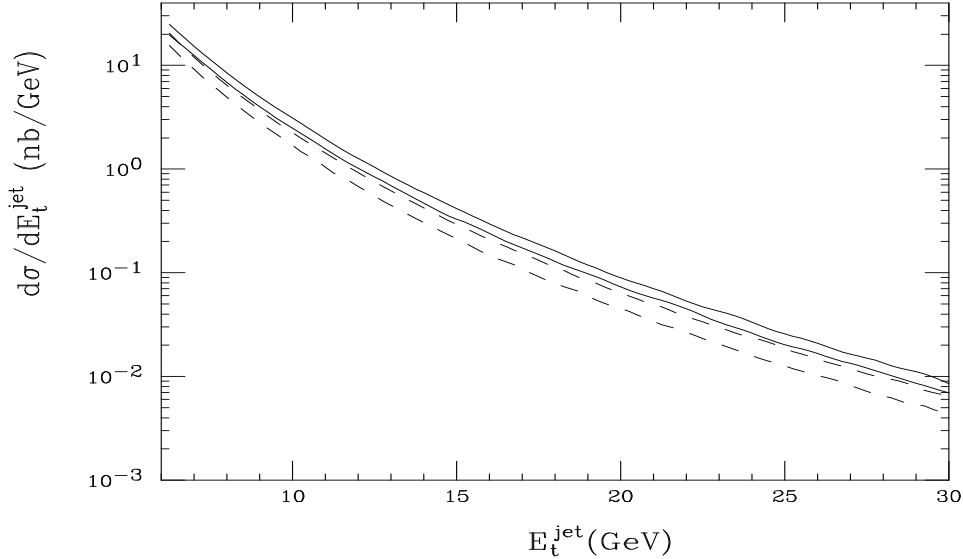,width=5.0in,height=3.09in}}}
\caption{Scale dependence of the single jet inclusive cross section as a 
function of $E_T^{{\rm jet}}$ integrated over $-1<\eta^{{\rm jet}}<2$.  
The LO (dash line) and NLO (solid line) results 
are shown for two different scale choices, $\mu=E_T^{{\rm max}}/2$ (top) 
and $\mu=2E_T^{{\rm max}}$ (bottom).}
\end{figure}
\begin{figure}
\centerline{\hbox{\psfig{figure=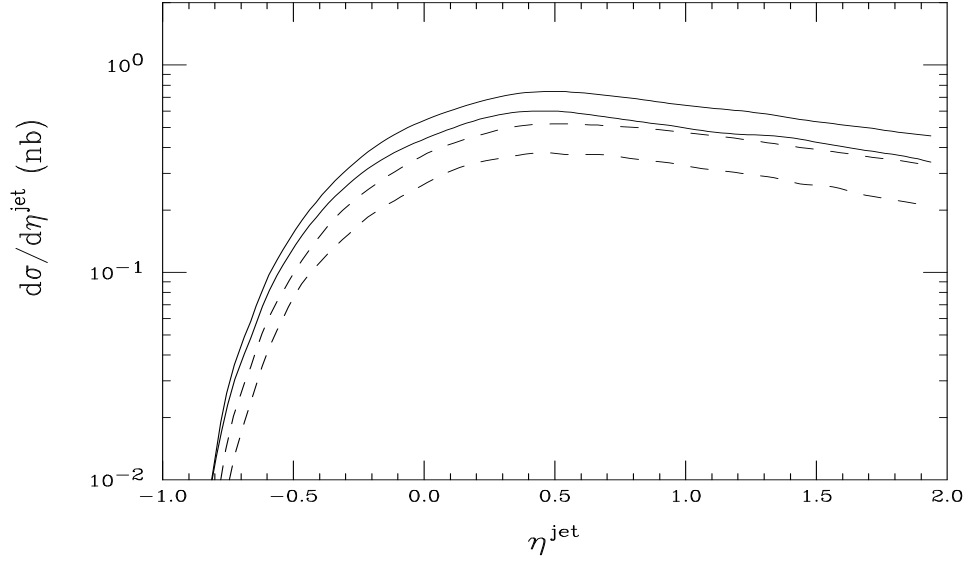,width=5.0in,height=3.09in}}}
\caption{Scale dependence of the single jet inclusive cross section as a 
function of $\eta^{{\rm jet}}$ integrated over 
$E_T^{{\rm jet}}>15\, {\rm GeV}$.  The LO (dash line) 
and NLO (solid line) results are shown for 
two different scale choices $\mu=E_T^{{\rm max}}/2$ (top) and 
$\mu=2E_T^{{\rm max}}$ (bottom).}
\end{figure}
One of the primary reasons for carrying out a NLO calculation is that of 
reducing factorization and renormalization scale dependence.   This is 
illustrated in FIG.\ 2 for the single jet inclusive cross section as a 
function of $E_T^{{\rm jet}}$ integrated over $-1<\eta^{{\rm jet}}<2$ 
for the 1993 H1 conditions with $R=1$ and $R_{\rm sep}=2$.  
The factorization and renormalization scales have been set equal to $\mu$.  
The LO (dash line) and NLO (solid line) results are shown for two different 
scale choices $\mu=E_T^{{\rm max}}/2$ (top) and $\mu=2E_T^{{\rm max}}$ 
(bottom) where $E_T^{\rm max}$ is the maximum transverse energy 
associated with the relevent two- or three-body weight.  The single jet 
inclusive cross section as a function of $\eta^{{\rm jet}}$ integrated over 
$E_T^{{\rm jet}}>15\, {\rm GeV}$ is shown in FIG. 3.  In both cases, the NLO 
calculation shows reduced scale dependence relative to the LO 
result.  

\begin{figure}
\centerline{\hbox{\psfig{figure=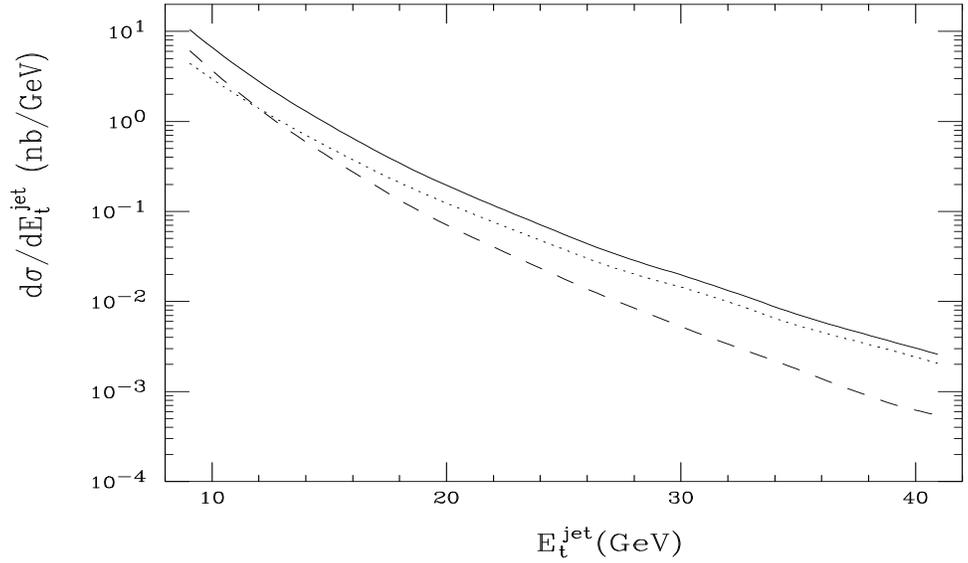,width=5.0in,height=3.09in}}}
\caption{The NLO single jet inclusive 
cross section as a function of $E_T^{{\rm jet}}$ integrated over 
$-1<\eta^{{\rm jet}}<2$ in three different regions $0<x_{\gamma}<1$ (solid), 
$x_{\gamma}>0.75$ (dot), and $x_{\gamma}<0.75$ (dash).}
\end{figure}
\begin{figure}
\centerline{\hbox{\psfig{figure=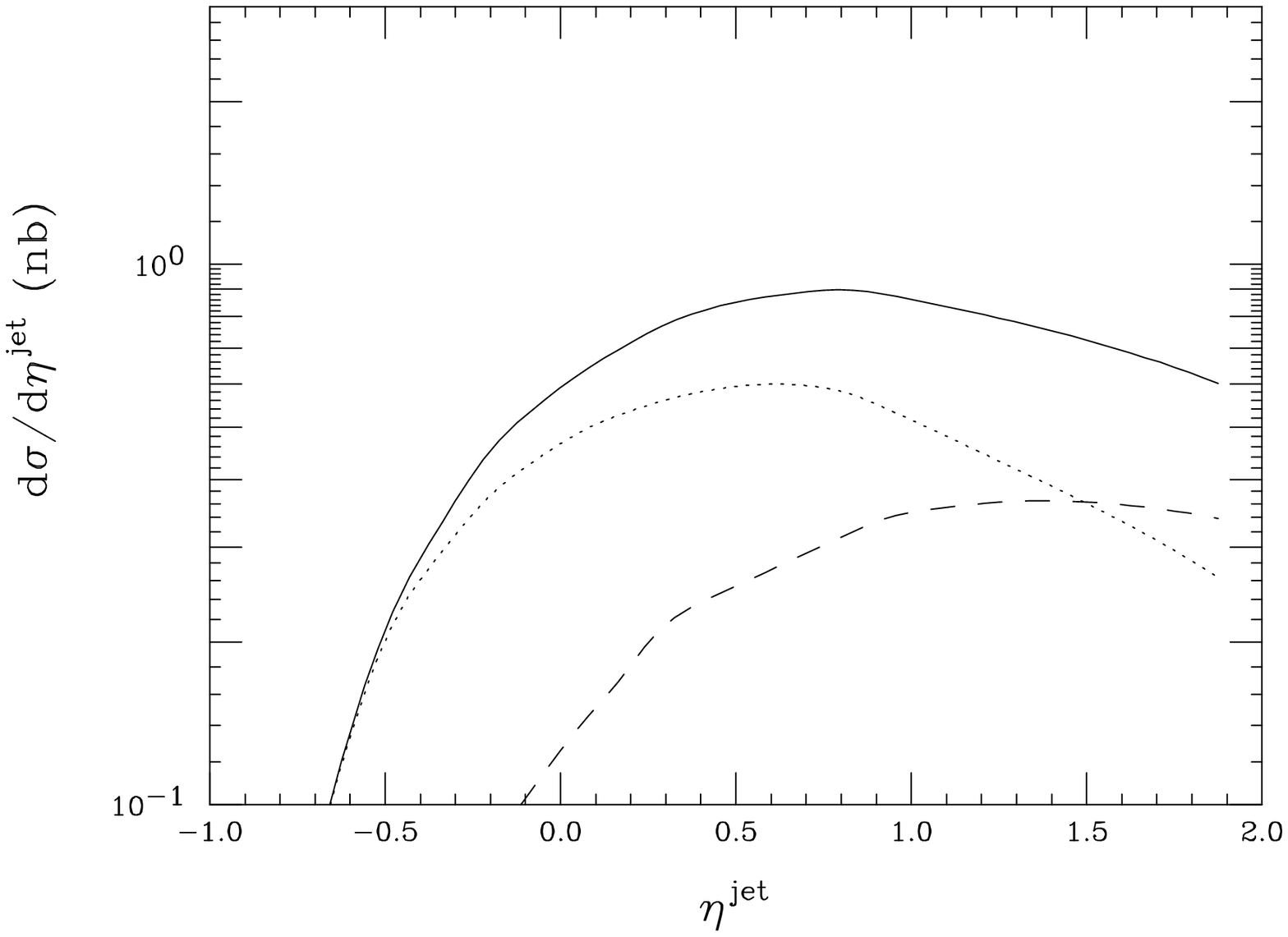,width=5.0in,height=3.09in}}}
\caption{The NLO single jet inclusive 
cross section as a function of $\eta^{{\rm jet}}$ integrated over 
$E_T^{{\rm jet}}>17$ in three different regions $0<x_{\gamma}<1$ (solid), 
$x_{\gamma}>0.75$ (dot), and $x_{\gamma}<0.75$ (dash).}
\end{figure}
\begin{figure}
\centerline{\hbox{\psfig{figure=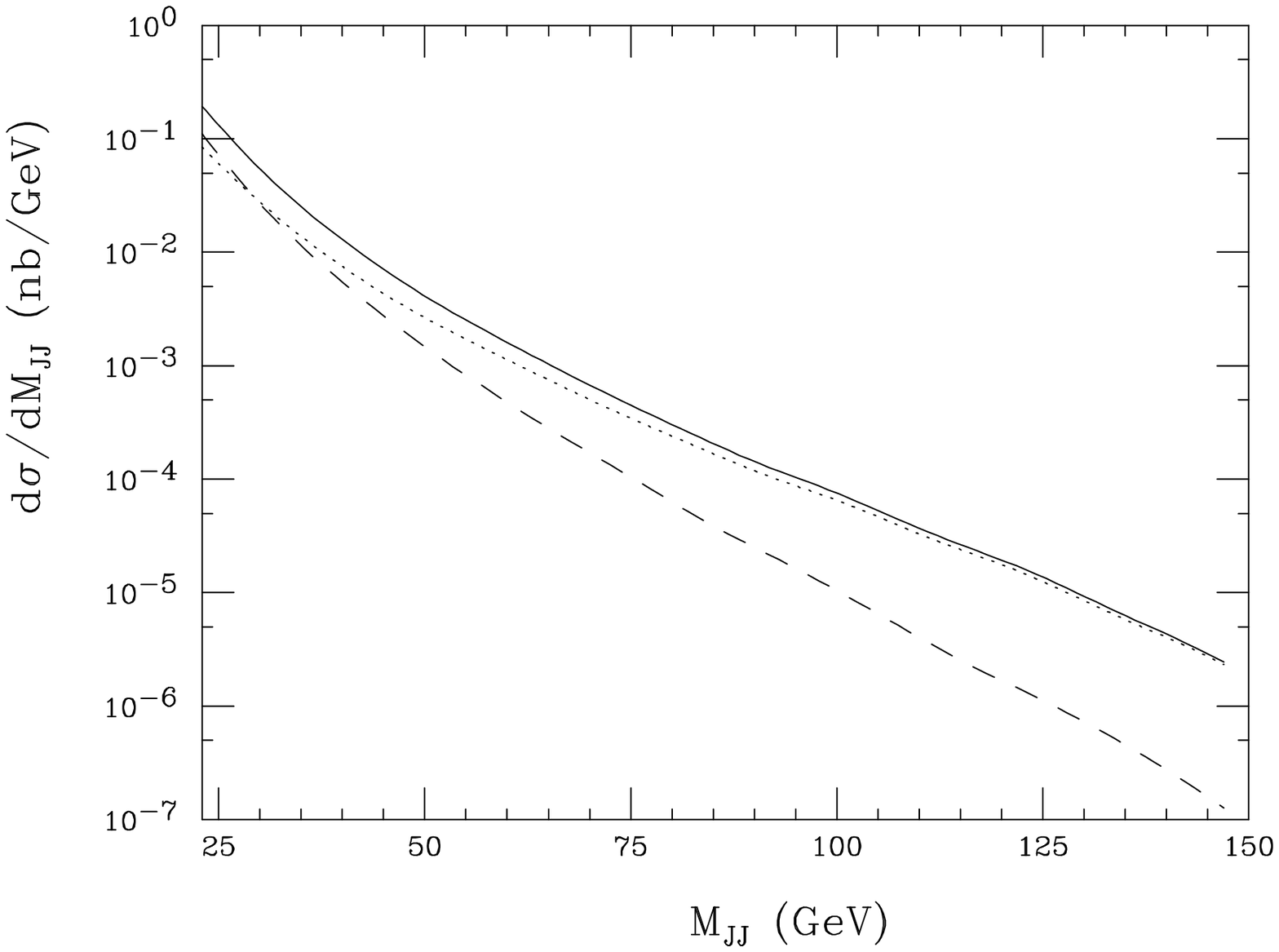,width=5.0in,height=3.09in}}}
\caption{Dijet cross section as a function of the invariant mass of the two 
highest $E_T$ jets decomposed into the regions $0<x_{\gamma}<1$ (solid), 
$x_{\gamma}>0.75$ (dot), and $x_{\gamma}<0.75$ (dash).}
\end{figure}

As stated in the introduction, it is conventional to describe jet 
photoproduction in terms of two components. The direct or point-like
component corresponds to the case where the photon participates wholly in the 
hard-scattering subprocess, whereas the resolved or hadron-like component
corresponds to a situation where the photon interacts as if it contained
partons.  In the direct component there is, then, no remnant beam jet and
this leads to a simpler event structure than in the purely hadronic case.
The decomposition is, however, an artifact of viewing the process from the
perspective of leading order QCD.  At next-to-leading order , the
two components mix and a complete separation is not possible.
 
Although a strict separation into direct and 
resolved components is not possible, one can still define kinematic 
regions where either component dominates.  One such definition \cite{xgamdef} 
uses the {\em observed} momentum fraction $x_{\gamma}$ of the parton coming 
from the photon which is given by  
\begin{equation}
x_{\gamma}=(E_T^{{\rm jet}_1} e^{-\eta_{{\rm \, jet}_1}} 
          + E_T^{{\rm jet}_2} e^{-\eta_{{\rm \, jet}_2}})/2E_{\gamma}.
\end{equation}
Here $E_T$ and $\eta$ are those of the two highest transverse energy jets.  
The region $x_{\gamma}>0.75$ corresponds primarily to the direct  
contribution and $x_{\gamma}<0.75$ to the resolved contribution.  

It is interesting to see how the two components populate phase space.  
To this end, shown in FIG.\ 4 is the NLO single jet inclusive cross section 
as a function of $E_T^{{\rm jet}}$ integrated over $-1<\eta^{{\rm jet}}<2$ 
in three different regions $0<x_{\gamma}<1$ (solid), $x_{\gamma}>0.75$ (dot), 
and $x_{\gamma}<0.75$ (dash) for the 1993 ZEUS conditions with 
$R=1$, and $R_{\rm sep}=2$.  
The cross section dominated by the resolved component has a steeper
$E_T$ dependence than that which is dominated by the direct component. 
This is due to the extra convolution associated with the photon parton 
distribution.  In other words, only a portion of the photon's energy 
contributes to the hard scattering for the resolved component.
Hence, at large $E_T$ values the direct component dominates.  
In FIG.\ 5 the NLO single jet inclusive cross section is shown as a function 
of $\eta^{{\rm jet}}$ integrated over $E_T^{{\rm jet}}>17\, {\rm GeV}$.  
The direct component dominates at negative rapidity whereas the 
resolved component takes over in the positive direction. In the coordinate
system used in the HERA experiments, positive rapidity corresponds to the
direction of the proton beam. Therefore, if the observed jet is in the positive
rapidity region, this corresponds to larger $x$ values from the proton and
smaller ones from the photon. Conversely, if the jet is in the negative
rapidity region, larger photon $x$ values and smaller proton $x$ values are
favored. Since the contribution dominated by the direct component has
$x_{\gamma} > 0.75$, it dominates in the negative rapidity region, as shown. 
In FIG.\ 6 the dijet cross section is shown as a function of the invariant 
mass of the two highest $E_T$ jets.  Again, the steeper resolved component 
dominates over the flatter direct component at low $M_{\rm JJ}$.

\begin{figure}
\centerline{\hbox{\psfig{figure=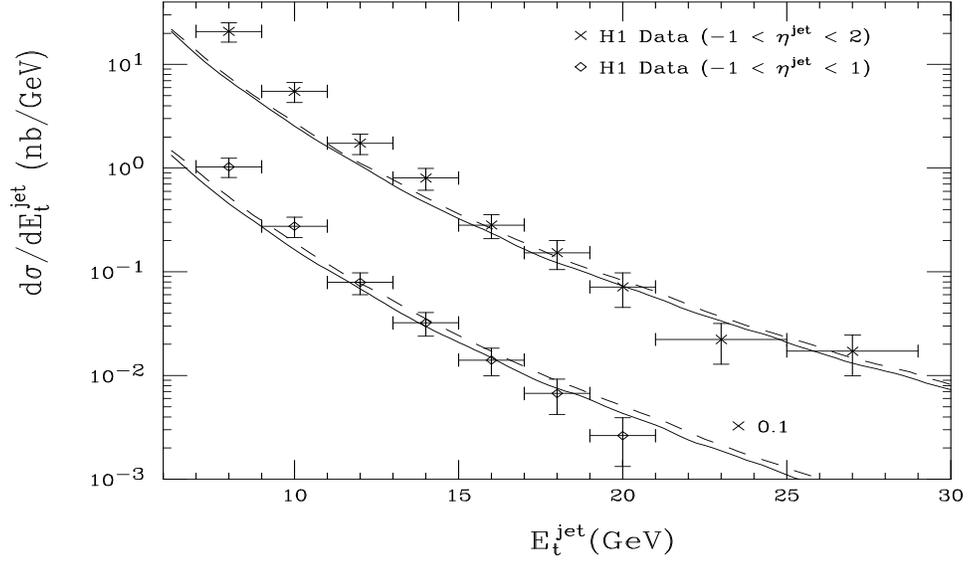,width=5.0in,height=3.09in}}}
\caption{The single jet inclusive cross section as a function of 
$E_T^{{\rm jet}}$ as measured by H1 \protect\cite{h1sngle} compared with 
our NLO result (GS photon: solid lines, GRV photon: dash lines).}
\end{figure}
\begin{figure}
\centerline{\hbox{\psfig{figure=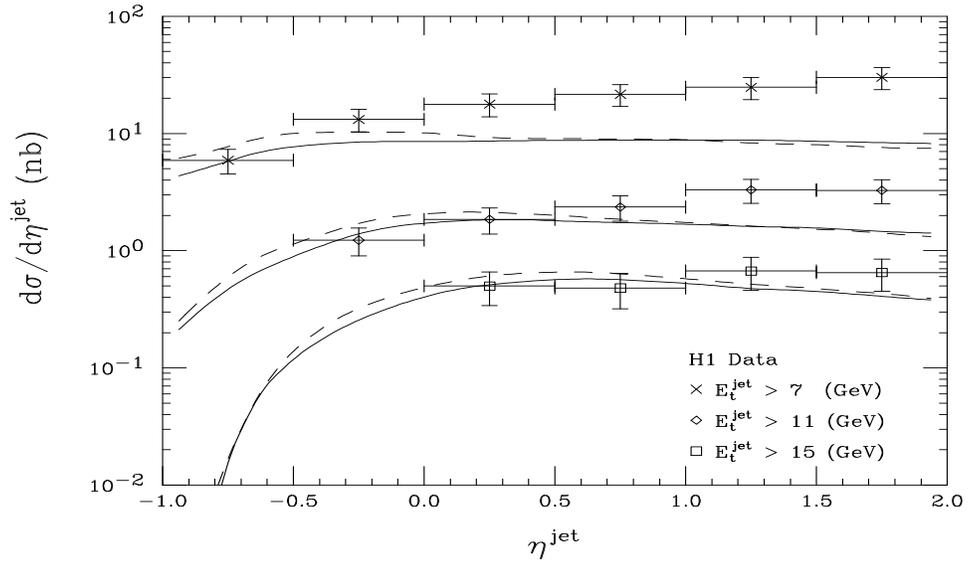,width=5.0in,height=3.09in}}}
\caption{The single jet inclusive cross section as a function of 
$\eta^{{\rm jet}}$ as measured by H1 \protect\cite{h1sngle} compared with 
our NLO result (GS photon: solid lines, GRV photon: dash lines).}
\end{figure}

\subsection{Comparison to single inclusive data}

Having discussed the general properties of the predictions, it is of interest
to see how they compare with existing data.
The first comparison is to the single jet inclusive cross sections as 
measured by H1 \cite{h1sngle}.  
The single jet inclusive cross section as a function of 
$E_T^{{\rm jet}}$ along with our NLO result 
(GS photon: solid lines, GRV photon: dash lines) 
for the 1993 H1 conditions with $\mu=E_T^{\rm max}$, $R=1$, 
and $R_{\rm sep}=2$ is shown in FIG.\ 7.
The data and theory have different slopes with the theory falling below the  
data at low $E_T^{\rm jet}$ for both photon sets.  
The agreement is slightly better for the 
case where $\eta^{jet}<1$.  
This is seen more clearly in FIG.\ 8 which shows the 
single jet inclusive cross section as a function of $\eta^{{\rm jet}}$ for 
$E_T^{\rm jet}>E_T^{\rm min}$ with 
$E_T^{\rm min}=7,11,{\rm and}\, 15\, {\rm GeV}$.  
In the proton direction the theory consistently undershoots the data.  
The agreement improves as $E_T^{\rm min}$ increases.  One possible 
explanation for the observed discrepency may be the need to increase 
the gluonic content of the photon which is poorly constrained 
in this $x$ region.  However, this would be premature as underlying 
event contributions, which were not removed from the data set or 
included in our calculation, may also provide an explaination.

\begin{figure}
\centerline{\hbox{\psfig{figure=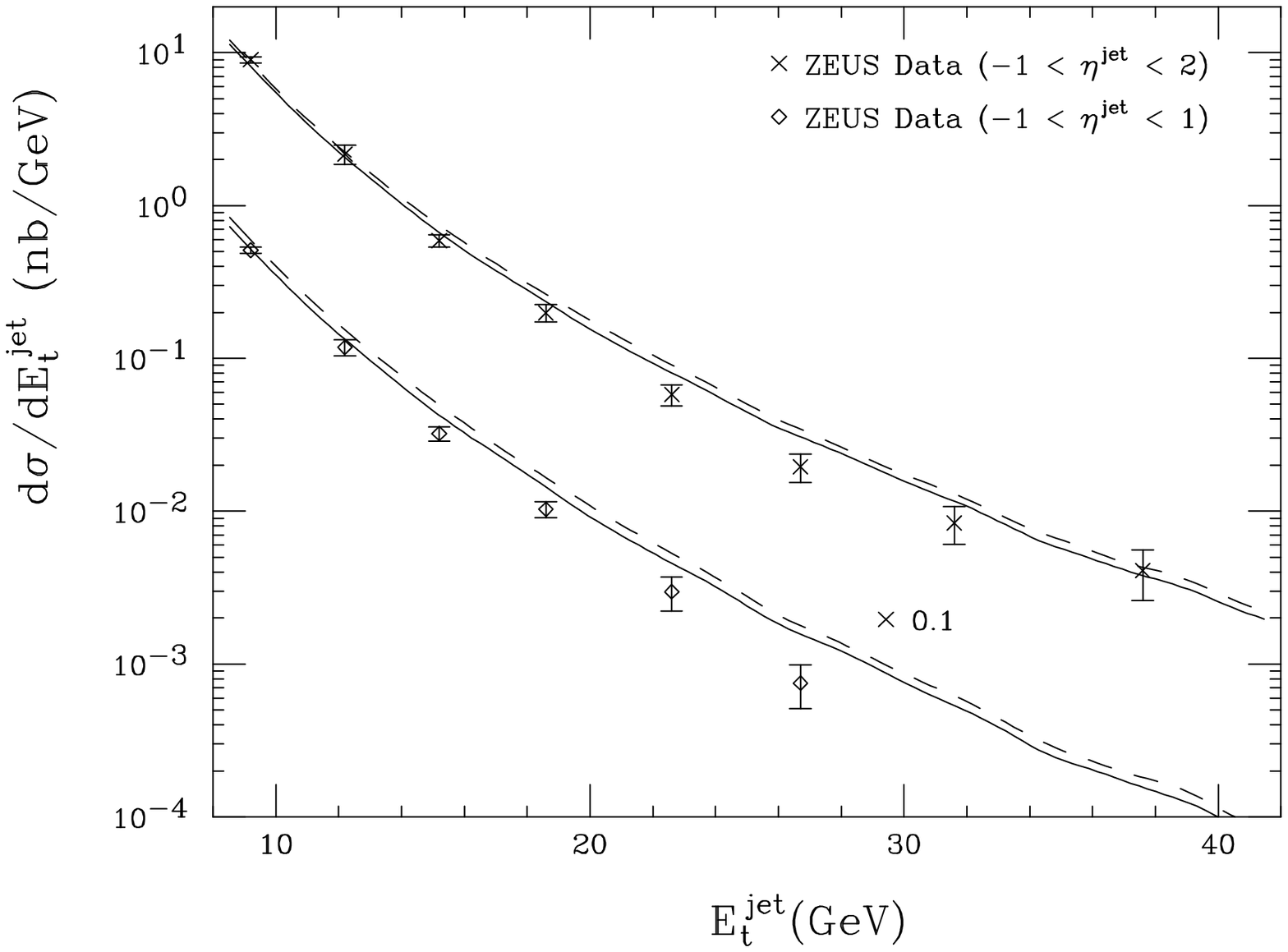,width=5.0in,height=3.09in}}}
\caption{The single jet inclusive cross section as a function of 
$E_T^{{\rm jet}}$ as measured by ZEUS \protect\cite{zeussngle} compared with 
our NLO result (GS photon: solid lines, GRV photon: dash lines).}
\end{figure}
\begin{figure}
\centerline{\hbox{\psfig{figure=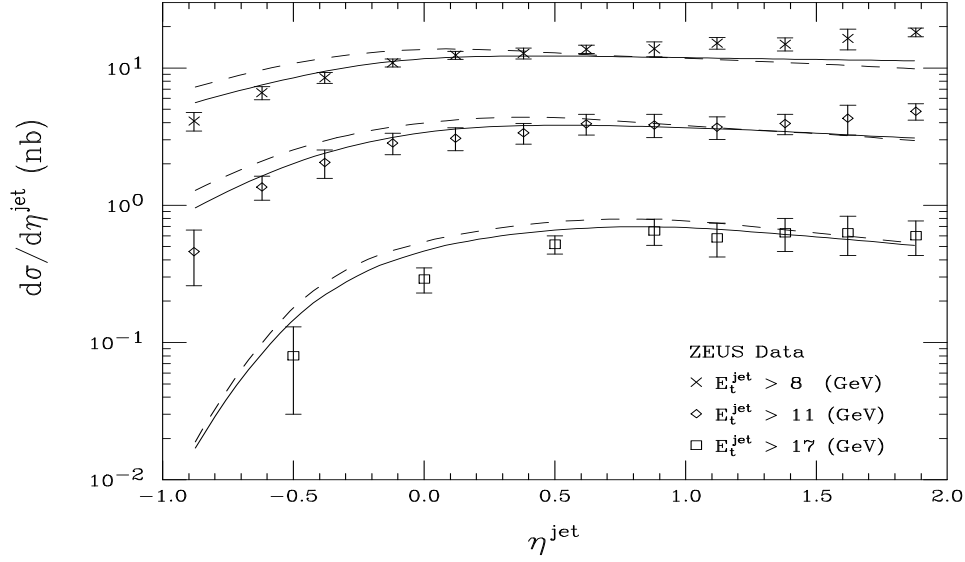,width=5.0in,height=3.09in}}}
\caption{The single jet inclusive cross section as a function of 
$\eta^{{\rm jet}}$ as measured by ZEUS \protect\cite{zeussngle} compared with 
our NLO result (GS photon: solid lines, GRV photon: dash lines).}
\end{figure}
We next switch to 1993 ZEUS conditions and compare to the published ZEUS 
data \cite{zeussngle} on single jet inclusive cross sections. In FIG.\ 9 
the measured single jet inclusive cross section is shown as a function of 
$E_T^{{\rm jet}}$ along with our NLO result 
(GS photon: solid lines, GRV photon: dash lines) with $\mu=E_T^{\rm max}$, 
$R=1$, and $R_{\rm sep}=2$.  In this case the agreement with the data is 
somewhat better than that shown in FIG. 7.   
This is further illustrated by the comparison with the single jet 
inclusive cross section as a function of $\eta^{{\rm jet}}$, shown in 
FIG.\ 10.  The trend, although less pronounced, is the same as with the H1 
data.

\subsection{Comparison to dijet data}

Data for several different types of observables involving two large transverse 
momentum jets have been published. In the ZEUS dijet analysis a sample of
dijet events is obtained by requiring 
$E_T^{{\rm jet}_1}, E_T^{{\rm jet}_2} > E_T^{\rm min}$.  
In principle, this 
presents a problem in the following sense.  The pair of jets can be accompanied
by additional soft gluons.  The singularities associated with the soft gluon
emission cancel infrared singularities coming from one-loop contributions.
This cancellation has already taken place in the present calculation.
That is, the soft gluons with energies up to $\delta_s \sqrt{s_{12}}/2$ have
already been integrated out. After the cancellation there are residual 
$\ln\delta_s$ terms in the two-body portion of the calculation which cancel
against logarithms built up by integrating over the unobserved gluon radiation 
in the three body contribution. The requirement that both jets have 
$E_T > E_T^{\rm min}$ restricts a portion of this integration and 
leads to an incomplete cancellation of the $\ln\delta_s$ dependence.  
We have, nevertheless, used this definition for the dijet 
sample, having found that the cutoff dependence of the results is 
negligible.  However, the problem still exists, at least in principle.

\begin{figure}
\centerline{\hbox{\psfig{figure=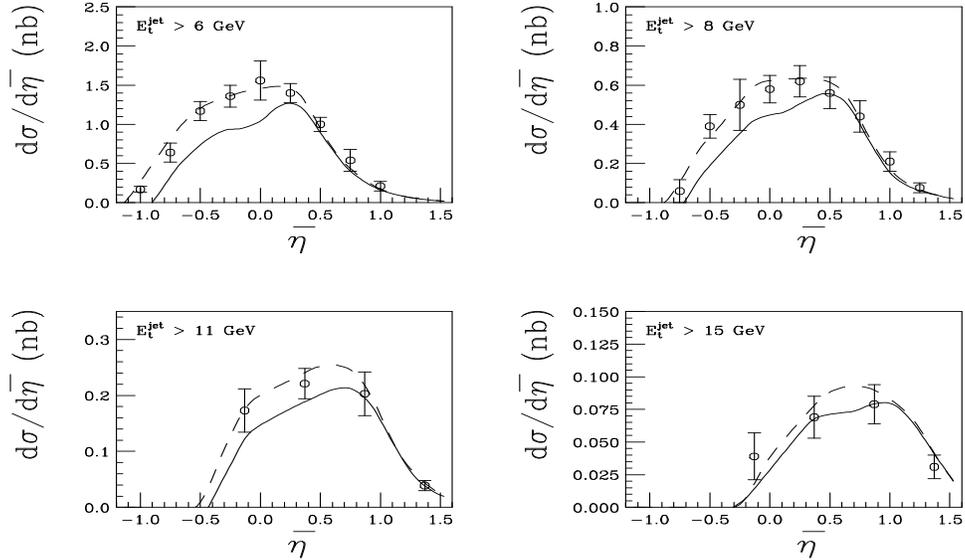,width=5.0in,height=3.09in}}}
\caption{Dijet cross section for $x_{\gamma}>0.75$ as a function of 
${\overline \eta}$ integrated over 
$E_T^{{\rm jet}}>E_T^{\rm min}$ for $E_T^{\rm min}=6,8,11,15\, {\rm GeV}$ as 
measured by ZEUS \protect\cite{zeusetabar}.  The solid curves correspond to 
requiring $E_T^{{\rm jet}_1}, E_T^{{\rm jet}_2} > E_T^{\rm min}$ while the 
dashed curves correspond to the cuts used in \protect\cite{kl}, as 
described in text.}
\end{figure}
\begin{figure}
\centerline{\hbox{\psfig{figure=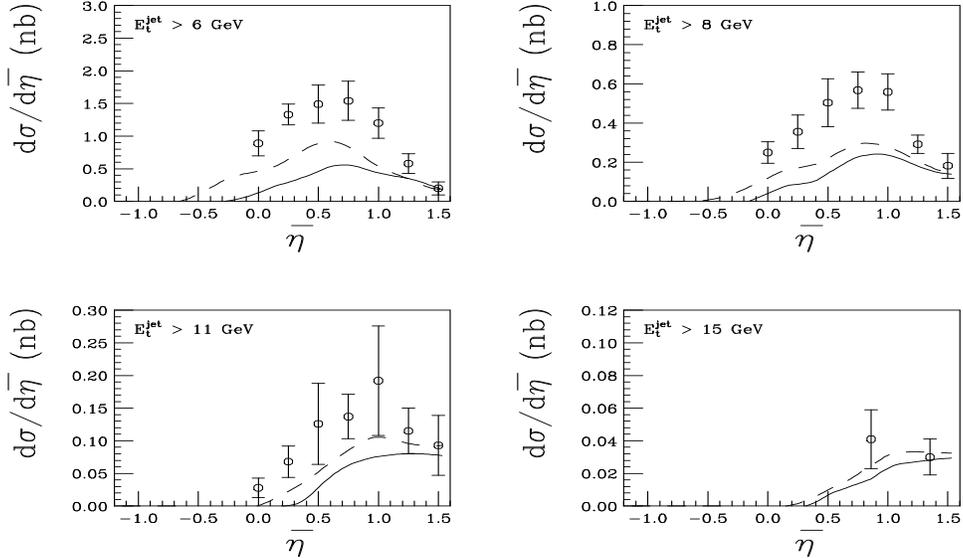,width=5.0in,height=3.09in}}}
\caption{Dijet cross section for $0.3<x_{\gamma}<0.75$ as a function of 
${\overline \eta}$ integrated over 
$E_T^{{\rm jet}}>E_T^{\rm min}$ for $E_T^{\rm min}=6,8,11,15\, {\rm GeV}$ as 
measured by ZEUS \protect\cite{zeusetabar}.  The solid curves correspond to 
requiring $E_T^{{\rm jet}_1}, E_T^{{\rm jet}_2} > E_T^{\rm min}$ while the 
dashed curves correspond to the cuts used in \protect\cite{kl}, as 
described in text.}
\end{figure}

Preliminary ZEUS data \cite{zeusetabar} for the dijet cross section as a 
function of ${\overline \eta}=(\eta^{{\rm jet}_1}+\eta^{{\rm jet}_2})/2$ 
integrated 
over $|\eta^{{\rm jet}_1}-\eta^{{\rm jet}_2}|<0.5$ are shown in 
FIGS.\ 11 and 12 for 
$E_T^{{\rm jet}_1},E_T^{{\rm jet}_2} > E_T^{\rm min}$ with 
$E_T^{\rm min}= 6,8,11,{\rm and}\, 15\, {\rm GeV}$. The solid curves 
are our results for the 1994 ZEUS conditions 
using the GRV photon set, $\mu=E_T^{\rm max}$, and $R=1$. A value of
$R_{sep}=1$ was used following \cite{workj} where it was found that
this value used at the parton level most closely corresponds to the KTCLUS
algorithm used to define the jets. No systematic cutoff dependence was observed
for these results, within the limitations of the Monte Carlo errors.
The solid curves display the same trend that was observed in the single
inclusive plots discussed previously insofar as 
the direct component comes closer to the data than does the resolved,  
the data are underestimated in the 
low $E_T$ region, and the agreement gets better as $E_T^{\rm min}$ increases.
This is consistent with there being some residual underlying event contribution
to the observed jets which has not been removed during the data analysis.

The problem of defining a sample of two jet events in a manner which is
amenable to a next-to-leading order theoretical treatment has been studied in 
\cite{kl}. 
In order to avoid a dependence on the theoretical cutoffs used in the course
of the calculation, it was proposed to calculate the dijet cross section by 
allowing the second jet to have 
a transverse energy less than $E_T^{\rm min}$ if the third unobserved 
jet is soft, i.e.\ has a transverse energy of less that $1\, {\rm GeV}$
($E_T^{{\rm jet}_3} < 1\, {\rm GeV}$). This approach slightly expands the
available phase space when both jets have transverse energies 
near $E_T^{\rm min}$ and,
therefore, avoids the problem of a potential cutoff dependence. Results using
this algorithm are shown in FIGS. 11 and 12 by the dashed lines. The relaxation
of the $E_T^{min}$ constraint on the second jet generates additional
contributions from the $2 \rightarrow 3$ subprocesses so that the dashed curves
lie above the solid ones and are in better agreement with the data. However,
the cuts do not correspond to those utilized in the definition of the data
sample. Given the unresolved question of the underlying event contributions, it
is premature to draw detailed conclusions from comparisons with these data.

\begin{figure}
\centerline{\hbox{\psfig{figure=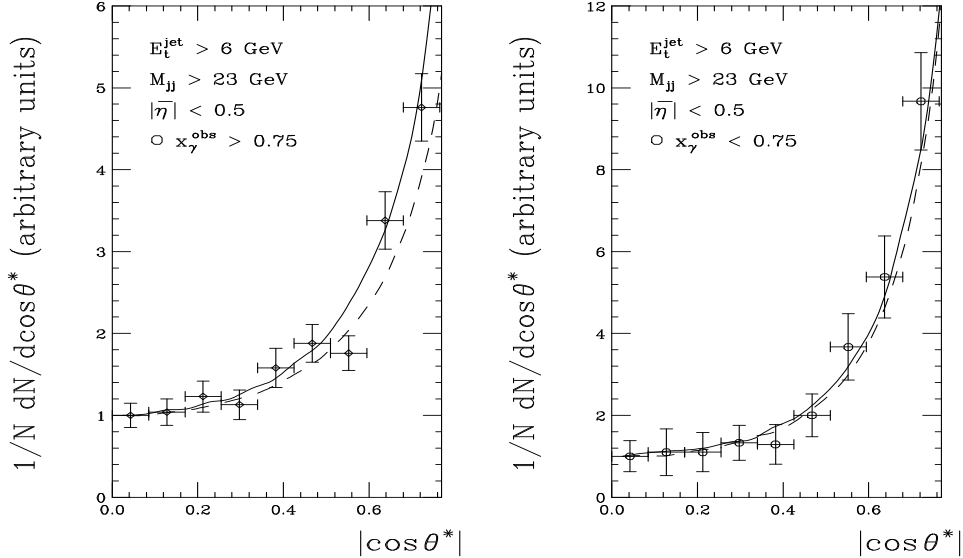,width=5.0in,height=3.09in}}}
\caption{Dijet angular distribution $d\sigma/d|\cos\theta^*|$ as measured by 
ZEUS \protect\cite{zeuscts} normalized to one at $\cos\theta^*=0$ compared 
with LO (dash lines) and NLO result 
(solid lines).}
\end{figure}

Angular distributions in dijet production are sensitive to the spin of 
the exchanged particle.  Let $\theta^*$ be the scattering angle of 
the two highest $E_T$ jets in their center of mass.  In LO, the 
direct component is dominated by $t-$channel quark exchange which 
has a characteristic angular dependence 
given by $(1-|\cos\theta^*|)^{-1}$.  The resolved 
component, however, is dominated by $t-$channel gluon exchange and 
has a $(1-|\cos\theta^*|)^{-2}$ angular dependence.
This property is preserved at NLO as demonstrated by the 
solid lines in FIG.\ 13 where the dijet angular distribution 
$d\sigma/d|\cos\theta^*|$ 
is shown as a function of $|\cos\theta^*|$ in the regions 
$x_{\gamma}>0.75$ (left) and $x_{\gamma}<0.75$ (right) 
for the 1994 ZEUS conditions with $R=1$, and $R_{\rm sep}=2$.
Also shown in the figure is the ZEUS \protect\cite{zeuscts} measurement 
and the LO (dash line) result.  
To better compare shapes, the distributions are normalized to one 
at $|\cos\theta^*|=0$.  The shapes are very insensitive to how 
one handles the dijet definition issue discussed above. The excellent agreement
between the predictions and the data confirms the theoretically expected 
behavior.

\section{Conclusion}

In this paper, results from a new next-to-leading-order 
program for jet photoproduction have been presented. The calculation was 
performed using the phase space slicing method and implemented 
in a Monte Carlo style program.  Physical observables are independent 
of the phase space slicing parameters.

As is the case in other jet production processes, the renormalization and 
factorization scale dependence 
is reduced relative to leading-order predictions.  
A decomposition into direct and resolved components was also 
studied using a definition based on the observed momentum fraction 
of the parton from the photon.  
The direct component dominates at large $E_T$ and in the 
photon direction while the resolved dominates at low $E_T$ and in 
the proton direction.

Comparisons with single inclusive jet cross section 
measurements by H1 and ZEUS as a function of $E_T^{\rm jet}$ and 
$\eta^{\rm jet}$ show that the theory tends to lie somewhat below the data in
the forward rapidity region. This is the region corresponding to low values
for $x_{\gamma}$ and which is dominated by the resolved component. One might
suspect that this indicates a need for increasing the size of the photon parton
distributions in this $x$ range, but this conclusion would be premature since 
contributions from underlying event structure could account for this,
as well.

Our results were also compared with dijet cross section measurements 
from ZEUS. The cuts used to define the dijet sample are such as to restrict a
portion of the phase space needed to insure the proper cancellation of the
cutoff dependence in the two- and three-body contributions. In practice, 
however, the variation was found to be smaller than the statistical errors
associated with the Monte Carlo integrations.
This point aside, the predictions for the normalized angular distributions 
agree well with the data. The ${\overline \eta}$ 
distributions show the same pattern of deviations from the theoretical
predictions as was observed in the single inclusive jet measurements.

As additional data are acquired it is possible that jet photoproduction will be
used to place additional constraints on the parton distributions in the photon.
Such data would complement data for the photon structure function which are 
primarily sensitive to the quark distributions in the photon. The jet
photoproduction process, on the other hand, directly involves the gluon
distribution in the photon. However, in order for such data to provide useful
constraints on the photon parton distributions, the effects of the underlying
event structure must be minimized. Higher $E_T$ thresholds combined with
a smaller cone size in the jet definition would help in this regard. 

%
%

\def\Journal#1#2#3#4{{#1} {\bf #2}, #3 (#4)}
\def\JCP{\em J. Comp. Phys.}
\def\JPG{{\em J. Phys.} G}
\def\NCA{\em Nuovo Cimento}
\def\NIM{\em Nucl. Instrum. Methods}
\def\NIMA{{\em Nucl. Instrum. Methods} A}
\def\NPB{{\em Nucl. Phys.} B}
\def\PLB{{\em Phys. Lett.}  B}
\def\PRL{\em Phys. Rev. Lett.}
\def\PRD{{\em Phys. Rev.} D}
\def\PR{\em Phys. Rev.}
\def\ZPC{{\em Z. Phys.} C}
\def\ZP{\em Z. Phys.}

\end{document}